\documentclass{ws-jai}
\usepackage[flushleft]{threeparttable}
\begin{document}

\catchline{}{}{}{}{} 

\markboth{P.Prasad}{The AARTFAAC All Sky Monitor: System Design and Implementation}

\title{The AARTFAAC All Sky Monitor: System Design and Implementation}

\author{Peeyush      Prasad$^\dagger$,      Folkert     Huizinga$^\S$,      Eric
  Kooistra$^\ddagger$,   Daniel  van   der  Schuur   $^\ddagger$,  Andre   Gunst
  $^\ddagger$, John Romein $^\ddagger$, Mark  Kuiack $^\S$, Gijs Molenaar $^\S$,
  Antonia Rowlinson $^\S$, John D. Swinbank $^\P$ and Ralph A.M.J Wijers$^\S$}

\address{ $^\dagger$Anton Pannekoek Institute,  University of Amsterdam, Postbus
  94249        1090        GE,         Amsterdam,        The        Netherlands,
  p.prasad@uva.nl\\   $^\ddagger$ASTRON,  Oude   Hoogeveensedijk,  7991PD,   The
  Netherlands\\ $^\S$Anton Pannekoek Institute, University of Amsterdam, Postbus
  94249 1090  GE, Amsterdam, The Netherlands\\  $^\P$Department of Astrophysical
  Sciences, Princeton University, Princeton, NJ 08544, USA\\ }

\maketitle

\corres{$^\dagger$Corresponding Author}

\begin{history}
\received{(to be inserted by publisher)};
\revised{(to be inserted by publisher)};
\accepted{(to be inserted by publisher)};
\end{history}

\begin{abstract}
The Amsterdam-ASTRON  Radio Transients  Facility And Analysis  Center (AARTFAAC)
all sky monitor  is a sensitive, real  time transient detector based  on the Low
Frequency Array  (LOFAR).  It generates  images of  the low frequency  radio sky
with spatial resolution of 10s of arcmin,  MHz bandwidths, and a time cadence of
a few seconds, while simultaneously but independently observing with LOFAR.  The
image timeseries  is then monitored  for short  and bright radio  transients. On
detection of  a transient, a  low latency trigger  will be generated  for LOFAR,
which can  interrupt its  schedule to  carry out  follow-up observations  of the
trigger  location  at high  sensitivity  and  resolutions.   In this  paper,  we
describe our heterogeneous, hierarchical design to manage the ~240 Gbps raw data
rate,  and  large scale  computing  to  produce  real-time images  with  minimum
latency.  We discuss the implementation of the instrumentation, its performance,
and scalability.
\end{abstract}


\section{\label{sec:Introduction}Introduction}

\noindent Transient astronomy  deals with the detection  and characterization of
celestial transients: sources in the  sky whose detectable properties can change
on short timescales.   These explosive events provide insight into  a variety of
astrophysics,  ranging from  emission mechanisms  of jets  to properties  of the
intervening  medium \citep{fender2006lofar,lazio2009dynamic,cordes2004dynamic}.  


The serendipitous discovery of a new  class of radio transient termed Fast Radio
Bursts  \citep[FRBs;][]{spitler2015fast, thornton2013population}  has galvanized
interest in the field.  The detected  FRBs are characterized by large associated
dispersion  measures,   high  brightness   and  short  timescales.    One  \cite
{spitler2016repeating}  discovered source  has been  found to  be non-repeating.
Their unknown origins makes it difficult  to discover such sources in a targeted
observation, with  only one possible  instance of a  multiwavelength association
with an FRB \cite {keane2016host}. Thus, discoveries during blind searches along
with a rapid follow-up over a  large wavelength regime are required to establish
their emission  phenomena and associated  parameters.  A recent example  of this
requirement is the detection by \citet{stewart2016lofar} of a ~20Jy transient in
60MHz Low Frequency Array (LOFAR)  data, whose characterization has suffered due
to inadequate multi-wavelength coverage.

For time-resolved observations, radio  instrumentation is generally available in
two  classes; Firstly,  a  single  dish or  phased  array  beam formed  approach
characterized by  high time and frequency  resolution, wide fields of  view of a
few degrees but poor spatial resolution. This mode is optimized for detection of
coherent   sources,   which  are   expected   to   emit  on   short   timescales
(milliseconds). Interferometric  aperture synthesis observations form  the other
class,  providing  high spatial  resolutions,  but  poor time  resolution.  They
typically need several  hours of observation time to build  up adequate coverage
in  the UV  plane  via earth  rotation aperture  synthesis  (however, see  \cite
{law2012rrat,  law2012all}).   This mode  is  optimal  for detecting  incoherent
sources, whose timescales of emission are much slower.

Large field of  view radio sky monitors are now  being developed to continuously
survey large parts of the visible sky  with shallow sensitivity and at high time
resolution,  in order  to  accelerate  transient discovery.   A  trigger can  be
generated on the  reliable detection of a transient in  near real-time, allowing
other telescopes to carry out follow-up observations.


The  Amsterdam-ASTRON Radio  Transient Facility  and Analysis  Center (AARTFAAC)
radio transient  monitor is such  an all-sky  radio transient detector.  It taps
data from a subset of the LOFAR antennas, and processes these data independently
of LOFAR.  It  is a leading effort  among a group of new  radio telescopes, with
other   notable    examples   being    the   Long   Wavelength    Array   (LWA),
\cite{ellingsonLWA1},   and  the   Murchison   Widefield   Array  (MWA),   \cite
     {tingay2013murchison}.   Such   telescopes  are  characterized   by  having
     moderate resolution and sensitivity as compared to contemporary telescopes,
     but  with  extremely  wide  fields   of  view  (typically  all  sky),  high
     availability, and autonomous calibration and imaging in near real time.

The AARTFAAC instrument also has secondary uses, besides its primary requirement
of  generating reliable  triggers.  Due  to the  time resolved,  wide field  and
continuous nature  of AARTFAAC  observations, its  secondary data  products like
all-sky images,  calibration solutions  and flagging information  find use  in a
variety of  science cases,  and for  LOFAR observatory  operations.  Application
areas include wide  field ionospheric monitoring via apparent  flux and position
variations of  calibrator sources, Solar  monitoring, RFI surveying,  LOFAR beam
model validation etc. The search for  fast transients across wide fields of view
will also be a  fundamental capability of phase 1 of  the Square Kilometer Array
(SKA) telescope \cite{colegate2011searching}.  The  AARTFAAC system is currently
the  largest   aperture  array   implementation  for  low   frequency  transient
monitoring. As such,  it provides a very realistic  test-bench for technological
approaches to all aspects of this kind of telescope.

The  wide field  of  views necessary  for  an instrument  like  AARTFAAC can  be
achieved by sampling the sky with wide-field dipoles. This, however comes at the
cost of lowered  sensitivity per receiving element.  A well  sampled UV plane is
needed  to  generate an  instantaneous  Point  Spread  Function (PSF)  with  low
sidelobes.  Both requirements  can be met by spatially spreading  a large number
of  dipoles.  However,  this requires  an order  of magnitude  larger number  of
elements in  the array than  contemporary arrays.  Bringing the  resulting large
number of data streams  to a central location, as well  as their correlation for
carrying out  aperture synthesis imaging in  real time thus poses  a significant
I/O and compute challenge. Further, the wide fields of view at the sensitivities
of operation also result in direction-dependent effects on the incoming signals,
mostly              due              to              the              ionosphere
\citep{intema2009ionospheric,wijnholds2010calibration}.  These  pose a challenge
to calibration, especially when carried out in an autonomous manner.


In  this paper,  we describe  the  AARTFAAC telescope  system architecture,  its
instrumentation,  and  the commissioning  of  its  various subsystems.   Section
\ref{sec:aartfaac_array} describes the array and the receiving antenna elements,
its  relationship  with LOFAR,  and  introduces  the  full architecture  of  the
instrument.    Section   \ref{sec:station_hardware}   describes   the   hardware
implementation in  the field which  allows creating a  data path in  parallel to
LOFAR. This  makes AARTFAAC processing independent  of LOFAR to a  large extent.
In Section \ref{sec:gpucorr}, we describe the implementation of a real-time, GPU
based  correlator  for  AARTFAAC,  while  Section  \ref{sec:calim}  details  the
real-time,  autonomous  calibration  and   imaging  implementation.  In  Section
\ref{sec:afaac_trap}, we  elaborate on the actual  transient detection mechanism
of the system.  Section \ref{sec:acontrol}  describes our control system for the
full instrument, which also interfaces with LOFAR.  In Section \ref{sec:results}
we present performance metrics of the instrument as a whole.

\section {\label{sec:aartfaac_array}The AARTFAAC Radio Transient Detection System}
We  begin by  summarizing the  subsystems of  the LOFAR  telescope relevant  for
AARTFAAC processing  in Section  \ref{subsec:lofar}, and  then elaborate  on the
scheme for creating a coupled data path for independent processing by AARTFAAC.

\subsection {\label{subsec:lofar} LOFAR Telescope Architecture}
The   LOFAR   telescope  \citep{van2013lofar}   is   a   new  generation   radio
interferometer  covering  the frequency  range  from  10-90 MHz  using  inverted
V-dipoles known as the Low-Band Antenna (LBA), and from 110-240 MHz using Bowtie
dipoles, also known  as the High-Band Antenna (HBA).  The  antennas are linearly
polarized, being made  up of orthogonally placed dipoles.  The  LBA dipole has a
sensitivity pattern with  a 6dB field of  view of about $120^o$  at 60MHz, while
the HBA dipoles first undergo an analog phasing within a 4x4 tile, which results
in a  field of view of  about $20^o$ at 140  MHz.  Due to this  restriction, the
AARTFAAC array utilizes only the LBA component of the telescope currently.

The  telescope itself  consists of  a  large collection  of antennas,  spatially
organized  into several  'stations',  each consisting  of  96 dual-pol  antennas
spread over a circle  of diameter ~60m.  Due to limited  hardware at stations in
the  core of  the array,  only 48  antennas belonging  to one  of a  few layouts
(including one  termed LBA\_OUTER) can be  utilized at any point.   The stations
are laid  out in a dense  core: 24 stations within  a 2km radius, with  the long
baselines made up using stations up to 1000km away from the core.

In the  regular LOFAR station  level processing,  the sampled bandwidth  of each
dipole  is split  into subbands,  which are  then digitally  phased in  hardware
towards the  direction of  an astronomical  source to form  a station  beam. The
phasing  is updated  periodically to  track the  position of  the source  in the
sky. The  resulting beam  is then  transmitted over optical  fiber to  a central
location for further interferometric processing with other stations.

\subsection {\label{subsec:aartfaac}  The AARTFAAC System Architecture}
\begin{figure*}[htbp]
\centering
\includegraphics[width=1\textwidth]{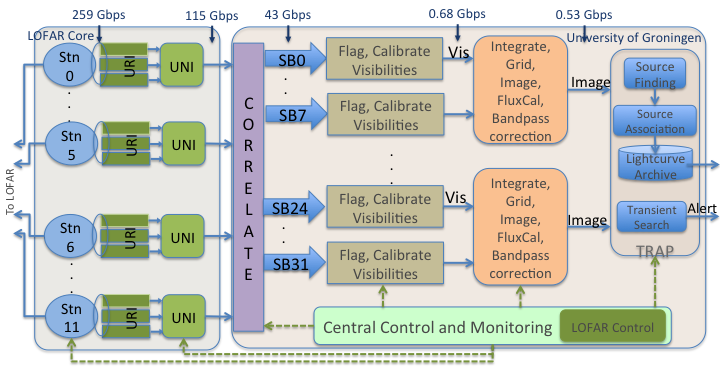}
\caption {Overall  architecture of the  AARTFAAC all-sky monitor  depicting each
  processing sub block, along with the  Monitoring and Control (MAC) system. The
  data  rates   correspond  to   the  total   bandwidth  across   the  specified
  interface. Stn refers to a single LOFAR station, SB refers to a single subband
  of 195312.5 Hz. 'Vis' refers to  visibilities. The green dashed paths indicate
  control flow, while the blue solid paths indicate data flow. }
\label{fig:afaac_arch}
\end{figure*}

The LOFAR station constitutes the first component of the radio sky monitor. This
is the  only sub-system  shared with  LOFAR.  The  AARTFAAC monitor  consists of
further  subsystems which  are  independent of  LOFAR  processing.  Its  overall
architecture   is  shown   schematically  in   Fig.  \ref{fig:afaac_arch},   and
illustrates the main processing sub-blocks of the instrument, including the data
routing and processing blocks, as well as the control and monitoring flow.


\subsubsection {Array Configuration}

\begin{figure*}[htbp]
\includegraphics[width=\textwidth]{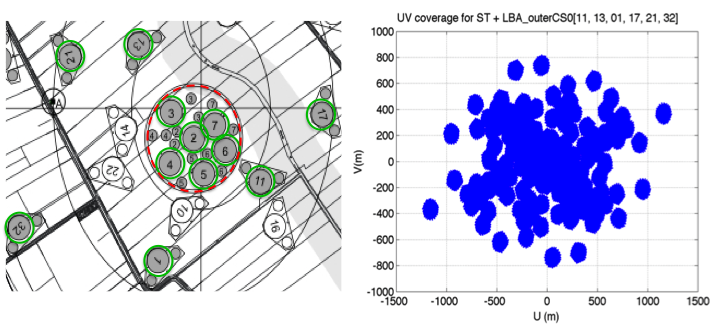}
\caption{(Left)The spatial distribution of  AARTFAAC-12 stations within the core
  of  LOFAR stations,  with LOFAR  station number  designation specified  in the
  green solid circles. The dashed red circle indicates the stations used for the
  AARTFAAC-6 subsystem. (Right) The instantaneous zenith pointing UV coverage of
  the AARTFAAC-12 system.}
\label{fig:afaac12_arrayconfig}
\end{figure*}

The  choice of  the subset  of LOFAR  stations used  in the  AARTFAAC system  is
dictated  primarily by  imaging  quality  and sensitivity,  as  well  as due  to
constraints       on       the       latency      of       calibration       and
imaging. Fig. \ref{fig:afaac12_arrayconfig} shows the stations in the LOFAR core
that are part of the AARTFAAC system, with inter-dipole distances ranging from a
few meters, to ~1236 meters. The central  six stations of the LOFAR telescope is
called  the  Superterp, and  is  indicated  by the  dashed  red  circle in  \ref
{fig:afaac12_arrayconfig}.  The Superterp forms a  densely sampled UV plane, and
is  ideal  for  wide field  imaging  since  it  is  co-planar to  high  accuracy
(centimeter level).   The outer  six stations  (circled in  solid green  in Fig.
\ref{fig:afaac12_arrayconfig})  provide higher  sensitivity and  resolution, and
have  been  chosen  as a  compromise  between  the  UV  coverage and  the  extra
processing  due to  the W-component.   The  salient features  of the  LBA\_OUTER
station   configuration   for  the   chosen   stations   are  shown   in   Table
\ref{tab:afaac_specs}.

In contrast to  LOFAR, AARTFAAC processes the data of  each individual dipole in
order to achieve all-sky imaging. Therefore a data spigot for each dipole signal
is created  prior to  the phasing up  of dipoles within  a LOFAR  station.  This
allows simultaneous  observing with  LOFAR, leading  to high  availability.  The
AARTFAAC system  could be built as  an off shoot  of LOFAR primarily due  to the
extreme  configurability offered  by LOFAR  system architecture.   However, this
restricts the layout  of the antennas within  a station as well  as the stations
themselves.  This results  in a  sub-optimal configuration  from the  transients
detection  perspective,  leading to  gaps  in  the instantaneous  12-station  UV
coverage,  and the  need  for  W-projection due  to  non-coplanarity across  the
stations.  Recognizing  this, the co-planar  AARTFAAC-6 system remains  a viable
sub-system for  certain science cases. It  consists of the 6  innermost stations
shown marked with a dashed red circle in Fig. \ref{fig:afaac12_arrayconfig}.

\subsubsection   {Hierarchical  Data   Shuffling  and  Transpose}
The high  bandwidth data  stream from  a large  number of  spatially distributed
dipoles  needs to  be ingested  while sustaining  the computationally  demanding
correlation operation. The  data also need arrangement in an  optimal manner for
the computing architecture, which amounts to a transpose operation.

We achieve both aims by spreading the collation and transpose operation over the
hardware on our data transmission  network. Intermediate nodes collect data from
different  input  streams and  exchange  their  dimensions, e.g.,  subbands  for
dipoles, by physically routing the data  out on different paths. Later, the same
operation  is carried  out in  the large  memories of  general purpose  CPUs via
sequential reads of data written in a strided manner to memory.  This delegation
of the transpose to various levels in the hierarchy is essential to managing the
large data rates, and to optimally use the computing infrastructure, and thus to
the functioning of such a telescope.


To summarize  the AARTFAAC processing, a  user selected subset of  subbands from
every  dipole is  transferred  as  UDP packets  over  a  dedicated 10Gbit  fiber
connection  to  the  central  processing  systems.   These  are  received  by  a
streaming, real-time  software correlator  implementation which aligns  the data
and estimates  the spatial covariance  matrix between  every pair of  dipoles at
high spectral and temporal resolutions.  The generated visibilities are streamed
over TCP/IP  to a calibration and  imaging pipeline component which  carries out
autonomous  imaging.  The  images  are  then analyzed  by  the LOFAR  Transients
Pipeline, \citep[TraP;][]{swinbank2015lofar}, which extracts the light curves of
sources within  the image, and analyses  them for variability using  a number of
parameters.   A (planned)  trigger  generation subsystem  will publish  reliable
triggers  in  the form  of  VOEvents  \cite{williams2006voevent}, which  can  be
claimed  by other  telescopes to  observe candidates  with high  sensitivity and
resolution.

\begin{wstable}[h]
\caption{Specifications of the AARTFAAC all-sky radio monitor.}
\begin{tabular}{@{}cccc@{}} \toprule
Parameter & Specification & Units & Comment\\ \colrule
Frequency range & 10-90 & MHz & Assuming LBA processing  \\
Processed bandwidth & 6.25 & MHz & Processing 32 subbands \\
Maximum baseline & 1236 & m & In LBA\_OUTER station array configuration\\
Resolution & 14 & arcmin & At 60 MHz \\
Sensitivity & 14\tnote{a} & Jy & 1 Subband, 1 sec integration \\
Frequency resolution & 1.56 & MHz & For transient detection. \\
 & & & Buffered visibilities at 3kHz resolution.\\
Time resolution & 1 & sec.\\ \colrule

\end{tabular}
\begin{tablenotes}
\item[a] Derived from AARTFAAC-6 measured sensitivity of 25 Jy.
\end{tablenotes}
\label{tab:afaac_specs}
\end{wstable}

We describe the various subsystems making up the AARTFAAC all-sky monitor in the
following sections.

\section {\label{sec:station_hardware} AARTFAAC Station Level Processing}
In  this section  the station  level instrumentation  relevant for  the AARTFAAC
system is discussed.  This involves systems which were already  present in LOFAR
stations, as well  as the additional instrumentation added  specifically for the
AARTFAAC system.
\subsection {Receivers}  The  antennas  in the  field  are connected  to
Receiver Units  (RCU). On those boards  the antenna signals are  amplified, band
pass limited  and converted into  the digital  domain.  The antenna  signals are
sampled with a 200 MHz clock frequency  delivering a bandwidth of 100 MHz to the
digital processing  system. The  A/D converter  uses a  sample resolution  of 12
bits.

\subsection  {LOFAR Digital  Processing}  The  LOFAR digital  processing
boards, also  referred as  Remote Station  Processing (RSP)  boards are  used to
channelize  the 100  MHz bandwidth  into 1024  subbands.  The  channelization is
implemented with a 1024-tap PolyPhase  filter bank implementation on each dipole
input. The output of  this filter bank is used for both  LOFAR and AARTFAAC. The
filter  bank analyzes  the sampled  voltage  timeseries into  a complex  voltage
spectrum  of 512  subbands. Thus,  the  entire analog  band of  the LBA  between
10-90MHz  is available  for further  processing,  of which  AARTFAAC utilizes  a
subset of about 6.25 MHz.

The output, for a set of 1024 real voltage samples, consists of a complex number
per  subband,  with a  2s  complement  16-bit  representation  of the  real  and
imaginary components.  A  single RSP board can handle the  processing of sampled
data from 4 dual polarized antennas. Since a LOFAR station is made up of 48 dual
polarized dipole antennas, 12 RSP boards are required per station. This is shown
in Fig.   \ref{fig:afaac_station_hw}.  For LOFAR  (and not for  AARTFAAC), these
subbands are  further processed  in the  RSP boards  to form  spatially directed
station  beams  by phasing  up  the  information  from  each antenna  for  every
subband. In  order to  combine the  information of all  antenna signals  the RSP
boards are connected via a ring  network. For LOFAR observations, each RSP board
is responsible to  calculate a partial beamformed sum of  the antennas connected
to that particular RSP board.  It further  adds this to the result received from
its preceding neighbour, and routes the partial sum onto the ring network to its
succeeding neighbour.  In this way the last RSP board in the ring calculates the
final beamformed sum, resulting in a spatially directed station beam.

The ring network consists of four 2  Gbps links, and is formed by daisy chaining
the serial I/O links of one RSP board to the next.  These links are also used to
carry AARTFAAC data  due to availability of left over  bandwidth.  Therefore, it
is now used to carry LOFAR specific data products (partial sums), along with the
raw AARTFAAC subbands.  Of the total  8Gbps bandwidth of the ring network, about
6 Gbps  is occupied  by LOFAR  specific products,  with the  remaining bandwidth
carrying per dipole subbands used exclusively for AARTFAAC processing.


The RSP firmware on  the FPGAs have been modified to  enable the transmission of
the  AARTFAAC data.   The AARTFAAC  specific firmware  selects a  subset of  the
available 512 subbands from all  antennas specifically and independent of LOFAR.
The available  ring network data  bandwidth forms the fundamental  limitation of
the  AARTFAAC processed  bandwidth.  To  retrieve more  bandwidth, the  firmware
accommodates a configuration in the number of bits used to represent the complex
filtered output per  subband.  Thus, bandwidth can be traded  with bit width, or
the dynamic range  in the filterbank outputs.   The bit mode of  AARTFAAC can be
set completely independently of LOFAR's choice  of bit mode.  The choice between
the various  bit modes depends  on the RFI  environment of the  observation.  An
8-bit complex representation of the filterbank subbands are found to be adequate
for almost all observing conditions except during severe RFI.

The bandwidth available to AARTFAAC is limited to 36 subbands in 16-bit mode, 72
subbands in  8-bit mode, 108 subbands  in 5 bit  mode, or 144 subbands  in 4-bit
mode.   This allows  AARTFAAC to  achieve high  sensitivity by  placing subbands
contiguously,  and later  integrating them,  while  at the  same time  achieving
spectral coverage  by placing subbands to  sample a larger extent  of the analog
spectrum. In the  rest of the document,  we refer to 8-bit  subbands.  Thus, the
current system  can generate 72  subbands, of  which 32 subbands  are processed,
corresponding to about 6.25 MHz (see \ref{sec:afaac_unb}).\\

\subsubsection  {Sampling Clock and  Timing}
 This sub-system is  shared with
LOFAR.  A  clock  distributor  board  (SyncOptics) at  the  center  is  used  to
distribute  a 10MHz  reference  to every  one  of the  24  core LOFAR  stations,
including the 12  AARTFAAC stations.  The 10MHz reference is  generated by a GPS
disciplined  Rubidium  frequency  standard,  and   is  fed  into  a  Timing  and
Distribution Board  at the  station.  This board  generates the  200MHz sampling
clock required by the RCUs, and is also  used for the data processing at the RSP
boards. It  ensures that  an identical  (hence coherent) clock  is used  for the
sampling of data from the AARTFAAC stations.

The absolute time is communicated to the  RSP boards on station reset by the LCU
(local control  unit) as a 64-bit  timestamp counter. The RSP  board then embeds
this 64-bit timestamp  into the data packets  that it generates at  the start of
the next absolute second, indicated by a  PPS (pulse per second) signal from the
GPS timing  system.  Once set,  the station hardware  updates this counter  on a
derivative of  the available 200MHz  reference, thus ensuring that  the absolute
time  is  embedded  in  the  data  with a  resolution  of  a  single  subband's
time-sample ($\sim5\mu sec$).  The absolute  timing accuracy depends on the long
term stability  of the  sampling clock  that is locked  to the  10MHz reference,
which in turn is derived from a GPS disciplined Rubidium frequency standard with
excellent long term stability.  All further  aligning and timing of the incoming
data in both  LOFAR and AARTFAAC systems  is carried out based  on this embedded
timestamp.

\subsection {AARTFAAC Piggyback System}
The AARTFAAC  piggyback system  is implemented  by URI  (UniBoard-RSP Interface)
boards.   These   boards   are   installed    in   the   ring   (as   shown   in
Fig. \ref{fig:afaac_station_hw}) and  basically tap-off the data  from each ring
interface. The ring  data is copied on  the URI board. One copy  is forwarded to
the next RSP  board in the ring  to carry out regular  LOFAR observations, while
the other copy is sent to the  AARTFAAC data router. Thus, the URI board ensures
high availability via simultaneous operation of LOFAR and AARTFAAC.

Each URI board  interfaces with the serial  I/O links of 4 RSP  boards.  The URI
board  further implements  the first  stage of  the overall  transpose operation
required  to bring  coincident  data of  all  dipoles for  a  single subband  to
consecutive  memory locations.   It  does  so by  statically  routing  up to  18
subbands from all dipoles available on the 4 RSP boards to a single output lane.
Each incoming link contains 72 subbands from 8 dipoles, while each outgoing link
contains  18  subbands  from  32  dipoles.    This  operation  can  be  seen  in
Fig. \ref{fig:afaac_station_hw} in the data-flow  layout between the URI and the
UNB Data  Router (explained  next), which  shows the collation  of data  from 18
subbands for 32  dipoles onto a single data link.   Altogether, three URI boards
are adequate to  transfer and transpose 72 subbands at  8 bits into the Uniboard
based router for all antennas within a station.\\

\subsubsection {\label {sec:afaac_unb} AARTFAAC  Data Router}
The data router  is the interface between the station  level instrumentation and
the next signal processing unit, the correlator.  The data router is implemented
with a UniBoard \citep{gunst2014application}.  The  board consists of 4 upstream
Field Programmable Gate  Arrays (FPGA) (called back-nodes) connected  to the URI
boards, and 4 downstream FPGAs (called front-nodes) to connect to the correlator
over  a  long  haul  fiber  link.   Each of  the  back-node  FPGAs  receives  18
consecutive subbands out  of the 72 subbands in the  URI board output.  However,
output bandwidth constraints from the  stations to the central signal processing
limit the  back-nodes to transferring  16 of the  18 subbands onward,  making 64
subbands available within this board.

A second level  of data rerouting is  carried out at this stage  such that links
from the  3 different URI boards  (each containing 18 subbands  from 32 dipoles)
are connected to  the same back-node.  This allows the  back-node to collect the
same subbands from  all 96 dipoles making  up the station, into  a single output
link.  The data from two back-nodes are transported to a single Front-node FPGA.
The latter  encapsulates the data  into a UDP packet  which is transmitted  on a
long  haul 10Gigabit  Ethernet interface  to  the remote  correlator.  

Due to limitations  of central processing, currently only  one front-node output
is utilized. Thus,  only 32 of the  64 subbands available from  the UniBoard are
transmitted to  the central  processing machines, located  at the  University of
Groningen about 50 km away.  Each station  output link carries about 9.7 Gbps of
data, consisting of 32 subbands of 8 bits from all dipoles in the station.\\



\subsubsection {Monitoring and Control  Interface}
Every station is equipped  with a Local Control Unit (LCU),  which is a computer
system running  a Linux operating  system.  These  systems are networked  to the
LOFAR  control system,  and also  act as  Network Time  Protocol (NTP)  clients.
Thus, their  absolute times are aligned  to better than a  few milliseconds. The
control  of the  remote station  electronics  consists of  two layers.  Firstly,
Command and Status Registers  have been opened up at the FPGA  level, and can be
accessed via a dedicated and separate  control Gigabit Ethernet interface to the
hardware boards.  Secondly, the  LCU provides an  abstraction layer  between the
hardware and  the global  LOFAR control  system via a  user accessible  tool and
driver combination.  All  control and monitoring commands from  a global control
system  are  addressed   to  the  LCU,  with  the   hardware  driver  ultimately
communicating the  commands over the  Gigabit Ethernet  control link to  the RSP
boards of the station.



\section {\label{sec:gpucorr} The AARTFAAC Real-Time Correlator}
\begin{figure*}[htbp]
\centering
\includegraphics[width=1\textwidth]{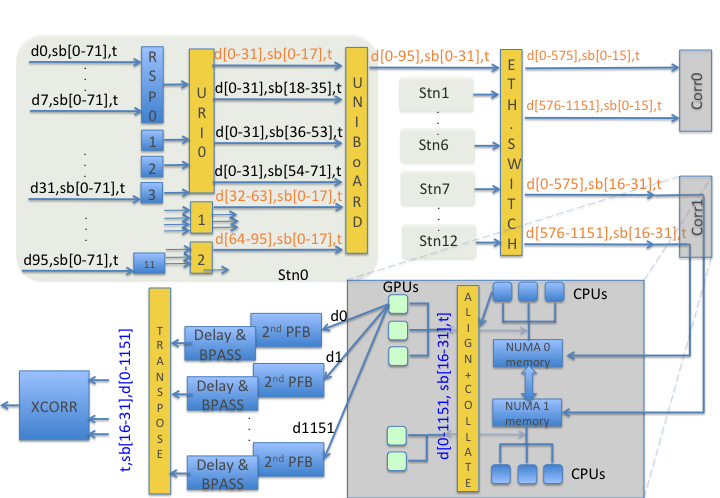}
\caption{The hierarchical  data processing and routing  necessary for optimizing
  correlator compute performance on GPUs.  Here, the flowing data is represented
  by the  triad of $[d[i-j],sb[i-j],t]$,  where $d[i-j]$ refer to  an individual
  dipole, $sb[i-j]$ refer to the range  of processed subbands, and $t$ refers to
  a time  sample.  The  yellow blocks correspond  to hierarchical  elements that
  shuffle data. $Stni$  refers to the data stream from  the ith station. $Corr0$
  and $Corr1$ are the two correlator machines.}
\label{fig:afaac_station_hw}
\end{figure*}

The correlator  subsystem estimates the  spatial coherence between all  pairs of
dipoles in the system per frequency channel, and time integrates the signal down
to 1 second.  With 12 stations, each containing 48  dual-polarized antennas, the
total number  of dipoles  is 1152,  making the correlator  one with  the highest
number of spatially distinct input streams among contemporary instruments.  This
is the most computationally intensive subsystem  of the pipeline, and the entire
data routing  hierarchy is fashioned  to lay  out the data  such that it  can be
optimally operated on by our chosen compute architecture.

The correlator's  input is  formed of the  subbanded complex  voltage timeseries
from each polarization of every antenna.  The output consists of a timeseries of
dipole array covariance matrices with a chosen time and frequency averaging.

The correlator ingests about 9.7 Gbps per station in real-time, corresponding to
32  subbands of  8 bits  for each  of  the 1152  dipoles.  It  needs to  produce
channelized data from the subband inputs due to the requirements of flagging and
calibration. The resulting computation is about 1 Tera Floating Point Operations
per  second (TFLOPs)  per subband  for  the full  array,  to be  carried out  in
real-time,  with  minimum   latency.   A  major  requirement   was  to  minimize
development effort, which effectively eliminated an FPGA based approach.

Of the available approaches, a  heterogeneous architecture consisting of general
purpose  CPUs in  combination with  Graphical Processing  Units (GPUs)  has been
found to be the  best match between our requirements of  ease of development and
performance.  Compared  to contemporary  multi-core CPUs and  DSP architectures,
GPUs have  been found  to have  the best performance  and energy  efficiency for
algorithms  relevant  to  channelizing  and  correlating  radio  astronomy  data
\cite{romein2016comparison}.  The  correlation operation  has a  high Arithmetic
Intensity, implying that  the data brought to a device  compute unit is operated
on many  times.  For a large  number of receivers, the  correlation operation is
thus compute bound. This implies that the  processing engines of the GPU are not
data starved due  to bandwidth limitations on  the PCIe bus between  the CPU and
the  GPU.   The  latest  server  class   machines  come  close  to  meeting  our
requirements of dense computing and high bandwidth I/O between CPUs and GPUs.

Our  correlator  implementation  shares  ancestry   with  the  LOFAR  GPU  based
correlator architecture,  which also  needs real time  processing to  reduce the
large  volumes of  data being  produced.  However,  LOFAR deals  with far  fewer
station input streams due to station  level beamforming, in turn processing many
more  subbands. This  results in  a very  different implementation  strategy for
both.


\subsection  {Implementation  Hardware  Architecture} 
The  heterogeneous AARTFAAC  correlator  is  made up  of  server class  machines
utilizing multiple  GPU devices to carry  out the computation necessary  for the
correlation.  The  host CPU acts as  the interface between the  station data and
the GPU devices.   They implement the data reception and  collation of data from
all stations, and arbitrate the data distribution between different GPUs.





The  implementation  consists  of  two identical  machine  configurations,  each
capable  of processing  16  incoming  subbands from  12  stations. Each  machine
consists  of  dual Xeon-class  processors  with  24  cores  each. The  CPUs  are
connected to  32 GB  of memory each,  operating in a  Non Uniform  Memory Access
(NUMA) configuration. The  bandwidth between these NUMA domains  is limited, and
minimizing the amount of memory  transfers between these domains complicated the
program code significantly.   Ten AMD FirePro S10000 dual GPU  cards (20 GPUs in
total) are available,  with 5 cards per machine.  These  interface with the CPUs
over PCIe3.0x16 lanes.  To receive the ~9.7 Gbps data output from each of the 12
stations, both  servers are equipped  with two 40Gbps Ethernet  interfaces, each
interface receiving  16 subbands from 6  stations, and also carrying  the output
correlations  to downstream  processors.  These  machines are  depicted in  Fig.
\ref{fig:afaac_station_hw} as Corr0 and Corr1.\\

\subsubsection  {High  Bandwidth  Switch}
An intermediate high bandwidth Ethernet switch, in combination with the Uniboard
data router,  is utilized to carry  out routing of  12 station data to  a single
machine.  The  Back-node FPGAs  of the  Uniboards encapsulate  data such  that a
subset of 16  subbands from 6 stations have an  identical destination IP address
of one of  the interfaces.  Thus, as shown  in Fig.  \ref{fig:afaac_station_hw},
half  the bandwidth  of stations  0-5 goes  to interface  0 of  Corr0, the  same
subbands of the stations 6-11 goes to  interface 1 of Corr0.  The remaining half
of the bandwidth goes in a similar fashion to the Corr1 machine.\\

\subsubsection  {NUMA  Domains}
Each correlator machine receives up to 16 subbands from all stations and manages
the  last-stage  data  exchange,  in   DRAM,  across  the  NUMA  domains.   (see
Fig. \ref{fig:afaac_station_hw}).   The resources available to  a single machine
are organized into  two NUMA domains on that machine,  with each domain handling
data from 6 stations. Each domain includes a 40Gbps Ethernet interface, a set of
processing cores, and the 32GB memory associated with them.

However, the split of  the 5 GPU cards per machine  cannot be made symmetrically
between the two domains, leaving one domain with 3 cards, while the other domain
as  2  cards.  The  NUMA  domains  allow binding  of  threads  handling I/O  and
processing to preferred CPU cores.  This prevents thread migration across cores,
which increases  data locality in  the core's  caches.  Thread binding  to cores
within a  NUMA domain allows routing  of network interrupts to  preferred cores.
This is essential for ensuring throughput of the system.

\subsection {Implementation of Functional  Blocks} 

In this  section, we provide a  description of the implementation  of individual
functional blocks on our target hardware.\\

\subsubsection  {CPU Data Collation and  Time Alignment}
The raw data input  packets from individual stations need to  be collated for an
entire integration period  in host memory, before being shipped  to the GPUs for
correlation.   4 of  the  12 available  processor  cores in  a  NUMA domain  are
dedicated for  handling the  Ethernet I/O interrupts. 

The other eight cores are used to  run the application threads.  By keeping four
cores free for interrupt handling, the  interrupt handlers do not need to switch
contexts  upon an  interrupt,  which  would be  too  time  consuming.  For  each
station, the  application starts an input  thread that receives the  station UDP
packets  and writes  the  data  into a  circular  buffer.   The circular  buffer
contains the last  four seconds of data, and is  continuously overwritten by new
data.  The main purpose of this buffer  is to time-align the station data before
further processing,  and to recover from  small hiccups in the  remainder of the
processing pipeline.

The correlator properly handles lost UDP packets from the stations, generating a
weighting  matrix  along  with  the integrated  visibilities.  The  aligned  and
collated subbands are then ready to be  transferred to a GPU’s global memory for
further processing.

\subsubsection {GPU Processing}
The GPU code is written in  OpenCL.  Every integration period (one second), when
the data from all stations should have  arrived, the host CPU starts pushing new
work to the  GPUs.  For each subband,  the host dynamically chooses  a free GPU.
Then, it  enqueues the data  transfer to the GPU,  all compute kernels,  and the
final transfer  of the visibilities  back to the  host.  The GPU  performs these
operations asynchronously.  Different subbands  are processed independently, and
are spread  over the available GPUs.   Below, we describe the  functional units;
more  details  about  the  GPU  code   implementation  can  be  found  in  \cite
{romein2016comparison}.\\



\subsubsection {Polyphase Filterbank}
 The first signal processing block is a
256 channel PolyPhase  filterbank. This consists of two kernels:  FIR filter and
FFT.   It is  applied onto  the  single subband  stream from  every dipole,  and
constituted of a bank  of 256, 16-tap FIR filters, followed by  a 1-D, 256 point
complex FFT  of the  filter outputs.   The FFT  is carried  out using  an openCL
library, while the  FIR filter is implemented  to maximize the usage  of the GPU
compute unit registers. The output is stored in the device memory.\\

\subsubsection {Delay,  Bandpass Compensation and Transpose  Kernel}
 A delay
compensation is applied  to the channelized data to account  for the fixed cable
delays of  the dipoles with  respect to a  reference antenna.  These  delays are
obtained via a separate calibration, which is typically carried out at a cadence
of  a few  months.  The  delays  are available  in calibration  tables, and  the
frequency resolution  is high  enough to  apply them as  phase rotations  of the
visibilities \cite {zatman1998narrow}.

The first stage polyphase filterbank implementation  in the RSP board results in
a deterministic  amplitude modulation  on the subband  bandpass, and  leading to
unequal powers in each subband channel.  This is demodulated via the application
of  a fixed  amplitude correction  by applying  channel-dependent weights.   The
resulting  data  is  brought  to   the  user  desired  frequency  resolution  by
integrating the channels.

Finally, the fine-grained parallelism axis  needs to be exchanged from frequency
channels to dipoles, which requires a transpose of the data.

\begin{figure*}[htbp]
\centering
\includegraphics[width=0.3\textwidth]{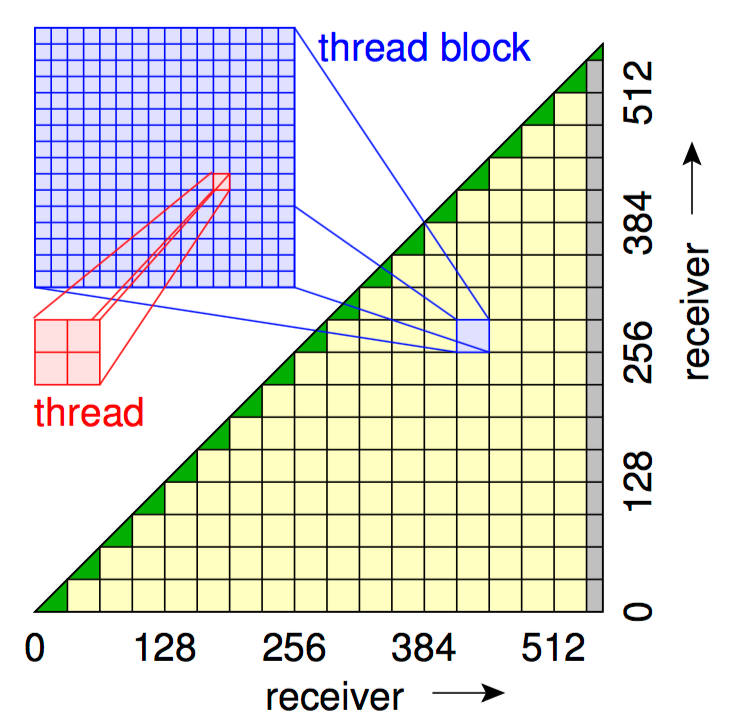}
\caption  {Depiction of  the splitting  of the  computed covariance  matrix into
  triangles, rectangles  and squares of outputs,  each unit being computed  by a
  compute unit of a GPU. From \cite{romein2016comparison}}
\label{fig:acm_spatial_split}
\end{figure*}

\subsubsection {Correlation Kernel}
Each  dipole's subbanded  and channelized  data is  then ready  for correlation.
Fig.  \ref{fig:acm_spatial_split} shows the output covariance matrix which needs
to be computed,  given the input per-dipole (mentioned  as receiver) channelized
data. Only one triangle of the hermitian covariance matrix needs to be computed,
as shown.

To distribute the computing of the correlations among the GPU compute units in a
given architecture, the covariance matrix  per frequency channel is divided into
squares and  triangles (with a  separate kernel to  evaluate each), as  shown in
Fig.   \ref{fig:acm_spatial_split}.  The  size  of the  squares  depends on  the
available  computing  and  memory  within  a  GPU  compute  unit,  and  for  our
architecture, squares of size 32x32 receivers is optimum. The computation of the
correlation of  all receivers  within a  square is  then a  \textit {work-group}
assigned to a GPU thread-block, with individual threads computing the covariance
of a 2x2 receiver tile within the  square of 32x32 receivers, as depicted.  This
arrangement results  in optimal  performance.  The actual  correlation operation
maps well to the efficient Fused Multiply Add (FMA) instructions of the GPU.  In
fact,  the  correlator  kernel  achieves  more  than  80\%  of  the  GPU's  peak
performance.   The resulting  visibilities  are kept  in  registers while  being
accumulated in time.   The integrated visibilities are written  to device global
memory by each GPU execution thread.  An event is generated on the completion of
the task.  The timing and weight  related meta-data is added to the correlations
to  form  the record  which  is  then streamed  out  over  a TCP  connection  to
downstream processors.\\

\subsubsection  {Asynchronous  Host  to Device  Transfers  Overlapping  with  Compute}
PCIe bandwidth is a scarce resource which we manage carefully. PCIe transfers to
host  memory  overlap well  with  computations,  to  avoid  GPU idle  times.  In
addition, we  avoid simultaneous  transfers between  the CPU  and GPUs  that are
connected through  the same  PCIe bus  (through PCIe switches  on board  the GPU
cards), so  that the  GPUs obtain  maximum bandwidth  and do  not have  to share
bandwidth  with other  GPUs on  the same  bus.  This  has been  found to  have a
profound effect on throughput and latency.\\

\section {\label{sec:calim} Real-time Flagging, Calibration and Imaging}
The real-time  flow of generated visibilities  need to be calibrated  and imaged
autonomously, and  with bounded latency.   This is a departure  from traditional
synthesis  imaging,  where the  long  observations  needed for  sensitivity  and
adequate   UV  coverage   are  bracketed   within  observations   of  calibrator
sources. The over-sampled instantaneous UV coverage,  the wide field of view and
the  relatively poor  instantaneous sensitivity  of the  AARTFAAC array  are the
reason  we use  a model  sky based  multi-source self  calibration approach,  as
described in more detail in \cite {prasad2014real}.

The calibration  and imaging  is carried  out on a  cluster of  multicore server
class machines, where each correlator output  subband is connected to a flagging
and calibration  pipeline, as  shown in Fig.   \ref{fig:afaac_arch}. We  use the
Eigen3 \citep{eigenweb} C++  template library to implement all  stages of matrix
processing and linear algebra in general within these pipelines.\\

\subsection{Calibration Pipeline}
Each channelized subband stream coming out  of the correlator is ingested into a
multi-threaded calibration pipeline with a ring buffer.  This ring buffer stores
the  raw visibilities  for approximately  $60$ seconds  in memory  which can  be
dumped to disk when triggered for  further examination by hand. A multi-consumer
lock-free queue grabs  the raw visibilities for processing  through the multiple
stages of  the calibration  pipeline. The processing  consists of  the following
operations.\\

\noindent \textbf {Weighting:} As the  correlator receives UDP data per station,
packets can get  lost. This information is sent to  the calibration pipeline and
corresponding visibilities are re-weighted to account for the loss of data. \\

\noindent \textbf  {Flagging:} The real  time flagging scheme consists  of sigma
clipping  the  visibilities based  on  their  amplitudes crossing  a  predefined
threshold above the  locally computed RMS. The clipping is  first applied across
visibilities  within   a  spectral  channel,  eliminating   those  crossing  the
threshold. This  is followed  by clipping channels  crossing the  threshold, for
every visibility.\\

\noindent  \textbf {Calibration:}  The  flagged visibilities  are amplitude  and
phase calibrated at the  channel level using a simple point  source model of the
four brightest sources  (Cas A, Cyg A, Vir  A, Tau A, termed the  A-team) in the
visible sky. The instantaneous shift of  source position from model locations is
estimated  using  the  Weighted Subspace  Fitting  \citep  {viberg1991detection}
algorithm. As part of the calibration process, the A-team sources are subtracted
out from  the calibrated  visibilities in  order to  reduce the  contribution of
their sidelobes  to the generated  images.  A  flux calibration is  then applied
based  on  the apparent  fluxes  of  the  A-team  during the  observation.   The
calibrated per channel visibility stream of  every subband is streamed out again
over  TCP  to a  machine  which  implements  the  actual spectral  and  temporal
integration to the desired level.\\

\subsection{Imaging Pipeline} 
The various calibrated  subbands streaming out of the  calibration pipelines are
merged  into  temporally  and  spectrally   integrated  images  in  the  imaging
pipeline. The processing consists of the following operations.\\

\noindent \textbf  {Visibility Gridding:} The  calibrated outputs of  8 subbands
are ingested  into a large  buffer and ordered  based on their  timestamps.  The
spectral integration is  carried out by gridding all visibilities  onto a common
grid  using   bi-linear  interpolation,  while  the   temporal  integration  (if
requested) is carried out by accumulation of the gridded visibilities.  Prior to
integration, another round of sigma clipping is carried out in both the spectral
and temporal axis to eliminate outliers.\\

\noindent   \textbf  {Imaging:}   The  multi-subband   integrated  and   gridded
visibilities are  Fourier Transformed to  generate the final snapshot  image. We
generate images of 1024 x 1024 pixels  for AARTFAAC-6, and of 2048 x 2048 pixels
for AARTFAAC-12 to adequately oversample the PSF. \\

\noindent \textbf {Beam model:} Finally, a beam model determined via simulations
of  the AARTFAAC  antennas is  applied in  the image  plane to  correct for  the
primary beam  response of  the dipoles  in the image.   The flux  calibration is
carried out using  measurements of the calibrator sources  established by \citet
{scaife2012broad}.  After  this stage,  the images  are sent  to TraP  (see next
section).

\section {\label{sec:afaac_trap} Transient Search Methodology}
The  Transients  Pipeline  \citep[TraP;][]{swinbank2015lofar}  carries  out  the
automated detection  of transients and  variable sources in AARTFAAC  images, in
near   real-time.     It   is    a   software    package\footnote{Available   at
  https://github.com/transientskp/tkp} optimized for the detection of transients
in radio images while specifically dealing with issues related to radio imaging,
e.g., noise correlation  or PSF related issues.  It consists  of a collection of
python processes carrying out image processing,  and a database which is used to
store the  image processing outputs  as well as to  carry out operations  on the
collective.   It operates  on  a  timeseries of  image  cubes  (each image  cube
consisting of two spatial and one spectral axis).

For any given image cube, it constructs  a catalog of all point sources (modeled
by elliptical Gaussians)  in every spectrally resolved image,  and compares them
against a database of point sources detected in previous timeslices.  The result
is the detection  of \textit{new} or \textit{variable} sources.   The former are
sources appearing  at locations where no  sources were seen in  previous epochs,
and the latter are sources which have been observed for multiple epochs and show
significant variability  in their  light curves  (timeseries of  detected source
intensities). These  results are  computed on the  basis of  the multi-frequency
light curves for every detected source, which are available in the database.

The TraP  is the  real-time consumer of  the generated  multi-frequency AARTFAAC
images,  and produces  two  outputs: A  trigger  to the  outside  world in  near
real-time on  the reliable detection  of a short-duration transient  or variable
source,  and  a spectrally  resolved  database  of  lightcurves of  all  sources
detected in a timeseries of image cubes, together with time-resolved information
about their variability.  The latter is available for both real-time and offline
data mining, e.g., to implement different transient detection approaches.

Two aspects of TraP are interesting from the AARTFAAC perspective: The effect of
AARTFAAC  specific characteristics  on  the image  processing,  and the  overall
latency  induced by  TraP operating  in a  streaming mode.  We discuss  these in
greater detail below.\\

\subsection  {Handling of  AARTFAAC  Specific  Characteristics by  TraP}
AARTFAAC creates instantaneous, transit  mode (non-tracking) all-sky images, and
will be continuously monitoring the sky. The  very wide field of view results in
a varying sensitivity  across an instantaneous image, which has  to be accounted
for before  islands of  high SNR  pixels can be  decomposed into  sources.  TraP
approaches this by modeling the background (mean) pixel value and RMS across the
image by  estimating these values  within every cell of  a grid laid  across the
image, and interpolating the values over the full image.

In spite of resolutions of a tens of arcmin, sources in AARTFAAC images can have
significant positional  jitter due  to ionospheric  effects.  TraP  accounts for
these during its \textit{source association}  step, when it identifies whether a
detected source can be associated with an existing source in its database, based
on spatial  proximity.  Due to  the non-tracking  nature of the  instrument, the
AARTFAAC sensitivity pattern is fixed  with respect to local coordinates. Hence,
point sources  can have very different  SNRs when they traverse  the sensitivity
pattern as they rise and set.  They can  thus be classified as a new source when
their SNR crosses detection thresholds.  TraP accommodates such cases by keeping
track of  the fields-of-view and sensitivities  of all images it  processes, and
comparing a  detected source's  flux density  against recorded  sensitivities of
images  covering  the  same  area  to  check if  it  could  have  been  detected
previously.

TraP  calculates  two  variability  metrics  for every  detected  source  in  an
instantaneous image, and  per frequency bin $\nu$: the  flux density coefficient
of variation $V_{\nu}$,  and the reduced weighted $\chi^2$ as  a significance of
flux  density  variability, denoted  by  $\eta_{\nu}$.  Although they  are  time
aggregated values based  on the lightcurve of the source,  they can be generated
iteratively  via running  statistics,  and are  available  for every  timeslice.
Thus,  the streaming  nature  of  AARTFAAC images  can  be  accommodated in  the
existing framework.

Since  theoretically  the AARTFAAC  image  stream  is infinite,  the  lightcurve
database needs to  be truncated in time.  Based on  the data-rates generated and
current  computing  capabilities,  a  database   containing  a  weeks  worth  of
lightcurves is manageable.   After this, a new database will  be created for the
next weeks'  observation. For requirements  of lightcurves with  duration longer
than a week,  wrapper scripts will be  used to query the  multiple databases and
construct required light curve.

Fig. \ref{fig:afaac_arch}  shows TraP as  the ultimate sink of  AARTFAAC images,
which will ingest 4 streams of image timeseries. Each stream corresponds to an 8
subband, 1 second integrated image timeseries.\\

\subsection {Latency}
TraP  latency   is  contributed   by  the  source   finder,  and   the  database
operation. The major operations of the  source finder are the RMS and background
map estimation, and fitting to detected sources. The former scales quadratically
with the number of pixels in an  individual image plane, while the latter scales
linearly with the  number of detected sources.  Database  operational times have
also been  shown to  scale linearly with  the number of  sources detected  in an
image. The total compute times of  the source finder and the database population
on AARTFAAC data have  been measured to be under a second  on test hardware (see
Table. \ref{tab:afaac_latency}).

\section {\label{sec:acontrol} The AARTFAAC Control System}
\begin{figure*}[htbp]
\centering
\includegraphics[width=0.8\textwidth]{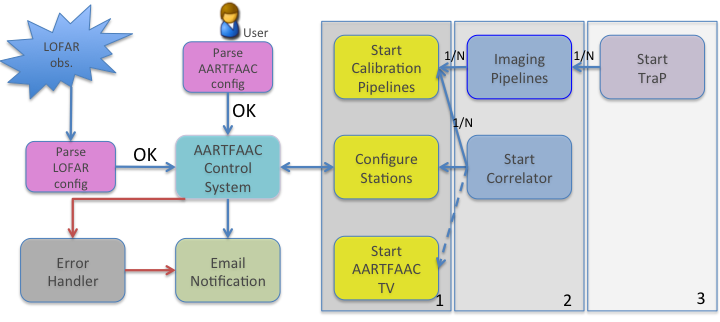}
\caption{The  control  system  architecture  which  interfaces  with  the  LOFAR
  observation  scheduling   system  and  triggers  AARTFAAC   observations.  The
  components are  organized into levels,  and the  direction of arrows  show the
  call-graph of dependency between the blocks.}
\label{fig:afaac_ctrl_sys}
\end{figure*}
The      AARTFAAC       control      subsystem       \footnote{Available      at
  https://github.com/transientskp/aartfaac-control}   coordinates  the   diverse
processing and I/O infrastructure of the  AARTFAAC system, and acts as a liaison
between the LOFAR observatory, AARTFAAC user and system.  It is essential to the
autonomous functioning of the instrument, and for providing fault tolerance.  It
has a python based client-server  architecture, with the server process existing
on  the LOFAR  manager node  and  clients waiting  for commands  on the  various
AARTFAAC  subsystem  controllers.    Fig.   \ref{fig:afaac_ctrl_sys}  shows  the
functional blocks  of the AARTFAAC  control system organized into  three levels,
indicated by the numbered blocks.

Every scheduled  LOFAR observation is  monitored for suitability as  an AARTFAAC
observation,  although  LOFAR  has  ultimate control  on  whether  AARTFAAC  can
piggyback on any  given observation.  When a LOFAR observation  is initiated and
AARTFAAC is  allowed to  piggyback, the control  system launches  the call-graph
using   the  current   active   AARTFAAC  configuration.    This  user   defined
configuration determines many  things, such as what subbands to  record, and how
many  images  to create.   The  call-graph  initiates  everything at  level  $1$
(stations, aartfaac tv, pipelines) by  connecting to the appropriate clients and
starting the processes.  When  at least $1$ out of $N$  pipelines is started and
the  stations  are  functioning  properly,  it calls  everything  at  level  $2$
(correlators and imaging  pipelines).  When at least $1$ out  of $N$ correlators
and $1$ out of $N$ imaging pipelines have been spawned, it will initiate TraP at
level  $3$.  stage.   An email  will be  sent when  an error  occurs or  when an
observation has successfully started.\\

\noindent \textbf  {Monitoring Interface:} The control  system allows monitoring
the  various  subsystems at  fine  granularity,  making  it useful  to  localize
problems within the system by examining  the email reports.  For monitoring data
flow         on         a          hardware         level         we         use
Munin\footnote{http://munin-monitoring.org}. This tool allows viewing statistics
of  I/O  between  nodes, computing  on  various  nodes,  and  disk usage  via  a
webpage\footnote{https://proxy.lofar.eu/aartfaac/munin},  including  history  at
various time cadences.

AARTFAAC TV is an application to  show live uncalibrated images, as an immediate
feedback of  the status  of an  observation.  The  processed outputs  at various
stages   in  the   pipeline  are   also  presented   by  AARTFAAC   TV  onto   a
webpage\footnote{https://proxy.lofar.eu/aartfaac/index.html} for the end to end,
and astronomical monitoring of the system.

\section {\label{sec:results} System Performance and Scalability}
\begin{figure*}[htbp]
   \includegraphics[width=\textwidth]{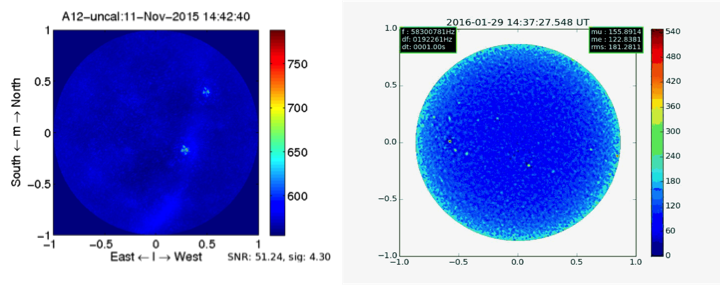}
   \caption{(Left)   An  uncalibrated   image   from   the  AARTFAAC-12   system
     demonstrating the hardware data routing and correlator functioning. (Right)
     A calibrated image from the AARTFAAC-6  system, with a bandwidth of ~1.5MHz
     and 1second integration.}
   \label{fig:afaac_snapshots}
 \end{figure*}


\begin{figure*}
  \includegraphics[width=\textwidth]{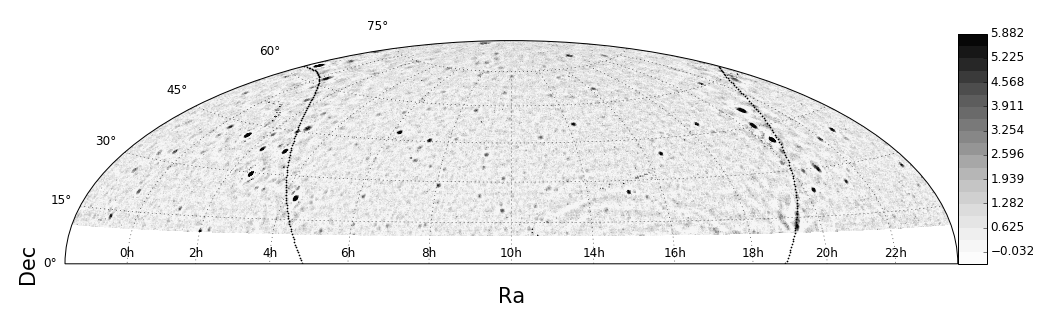}
  \caption{A calibrated sky-map created from a 24hr observation from AARTFAAC-6.}
  \label{fig:afaac_24hr}
  
\end{figure*}

Most components of the AARTFAAC-12 system have been commissioned. The correlator
has been tested  to operate at the  full specification and ~6.25
MHz  bandwidth.  The  12-station   calibration  sub-system  is  currently  being
commissioned,  Fig. \ref{fig:afaac_snapshots}  shows an  uncalibrated 12-station
image on the left.

The  AARTFAAC system  is  currently operating  with 6  LOFAR  stations, a  total
bandwidth of ~6.25 MHz, with real-time images  being created at a 1 second and 1
subband (195.3  kHz) cadence. The remaining  results and images in  this section
are  from  the  6-station  system.  Fig.   \ref{fig:afaac_snapshots}  shows  the
amplitude and phase calibrated all sky image from AARTFAAC-6 on the right, while
Fig.   \ref{fig:afaac_24hr}  shows  a  mosaic  of  the  whole  sky  as  seen  by
AARTFAAC. The  Galactic plane is  indicated by the  dashed line. The  latter has
been created  by combining  24 hours  of observations  taken at  1 second  and 1
subband integration.  The flux scale on all images is arbitrary. \\

\begin{wstable}[h]
\caption{Overall latency budget and performance of AARTFAAC subsystems.}
\begin{tabular}{@{}cccccc@{}} \toprule
 & A-12 & & A-6 & & \\ \colrule
Parameter & Compute & Processing  & Compute  & Processing  & Comment \\
 & time (ms) & TFLOPs & time (ms) & TFLOPs & \\
 \colrule
GPU FIR & 31.1 & 0.466 & 13.2 & 0.547 \\
GPU FFT & 18.6 & 0.488 & 9.18 & 0.493  \\
GPU Delay, BP, Trans. & 83.4 & 0.0054 & 25.7 & 0.0088  \\
GPU correlation & 289 & 2.98 & 69 & 2.12 \\
Online flagging & 199\tnote{a} & X  & 74.3 & X & \\
Calibration & 900\tnote{b} & X & 249 & X &  XX, YY pols, calibrating 63 channels.\\
 \colrule
Total & 1521 & & 440.3 & & \textbf{Measured AARTFAAC-6 latency} \\
 & & & & & \textbf {(upto calibration): 1100ms} \\ \colrule

Imaging & X & X  & 40 & & Gridding and FFT, 1 Stokes-I image.\\ 
TraP & X & X & 696 & &  Measured with 4 input subbands.\\ \colrule
\end{tabular}
\begin{tablenotes}
\item[a] Based on simulations.
\item[b] Based on extrapolating 6-station measurements.
\end{tablenotes}
\label{tab:afaac_latency}
\end{wstable}

\subsection {System Performance}
The overall  performance of the real-time  system is quantified by  the achieved
latency. Table  \ref{tab:afaac_latency} presents  the measured compute  time for
various functional blocks  of the system. Extrapolations on  the values measured
from the 6-station system are presented  for the calibration block, although the
presence of the  non-coplanar component would probably make this  a lower bound.
All reported  times have  been measured  on production  systems, except  for the
TraP.

Here,  we  see  that  the  most   compute  intensive  functional  block  is  the
calibration. Its compute footprint is dominated by the Weighted Subspace Fitting
model source position determination sub-block, and scales quadratically with the
number of input dipoles. Alternative approaches to this algorithm implementation
will  be  explored  to  reduce  this   cost,  and  thus  latency.   The  current
implementation  is adequate  to  maintain a  real  time throughput.  Calibration
latency can also  vary based on the observing conditions  like RFI occupancy and
the presence of the  flaring Sun. We thus set an upper  limit to the calibration
iterations, trading off instrumental sensitivity to maintain latency.

The next  compute intensive component is  the correlator itself. Its  latency is
independent of missing  or poor quality data, since the  collated data buffer is
processed based on wall-clock time. Thread binding to CPU cores prevents process
migration,  and  the absence  of  competing  processes reduce  operating  system
induced  non-deterministic  latencies.  We  see   that  the  FIR,  FFT  and  the
delay,bandpass correction and transpose block  scale linearly with the number of
dipoles, as expected. The correlation scales almost quadratically with input set
size.   In  our  measurements,  the   correlator  achieves  a  71\%  operational
efficiency of the theoretical maximum.

For the  6-station system,  we measure  a latency of  ~1.1 second  to calibrated
image generation,  on production hardware.  This value is obtained  by comparing
the  hardware generated  timestamp at  the  end of  the correlation  integration
period, to the wallclock time on  completion of calibration. This is about twice
as expected  from timings of the  compute blocks.  The remaining  latency budget
includes unmeasured,  but significant latencies.   These are caused by  the wide
area network  (10s of ms), correlator  input buffering, and host  to device I/O,
among other factors.

We expect  the total latency to  grow by factors  of 2-3 in anticipation  of the
more complicated calibration and imaging scheme for the 12-station AARTFAAC.

A streaming variant  of the TraP is still being  commissioned. Current profiling
reveals that  the source finding step  takes ~30\% of user  time, while database
operations to update the light curves  of detected sources take about 15\%. Both
these operations scale linearly with the number of detected sources in an image,
and thus  latencies on the  12-station images are  expected to be  only slightly
more.\\

\subsection   {AARTFAAC  Scalability}
The AARTFAAC all-sky  monitor implementation can be scaled up  along the spatial
(number  of dipoles)  or spectral  (processed subbands)  dimensions. A  spectral
scaling will  ultimately be limited  by the ring  network bandwidth to  about 64
subbands ($\sim12.5  MHz$) of 8 bits,  doubling the current bandwidth.   A spare
10Gbps link  on the  uniboards can bring  the extra 32  subbands to  the center,
where they  can be  processed by  an exact duplicate  of the  current correlator
system.   The choice  of  a  hierarchical data  transpose  results  in the  most
efficient final layout for correlation.  In the generic case, following the same
design, additional  subbands could be  accommodated by increasing the  levels in
the network hierarchy to accommodate the transpose.  The final correlation would
then be  possible for the  additional subbands  by replication of  the frequency
multiplexed hybrid correlator.

Keeping the  current bandwidth while increasing  the number of input  dipoles is
also feasible. The current Ethernet interface  bandwidths on the server side can
allow  another two  stations to  be added.   The compute  requirements would  be
almost 30\% higher, due to their  quadratic growth with number of input streams.
The correlation operation has been tested on the current GPUs for scalability of
the input streams, and should be able to cope with the requirements of the extra
inputs \cite {romein2016comparison}.  Accommodating  stations in addition to the
two  mentioned  here  would  require  additional  correlator  machines  and  the
transpose to be performed over a high  speed network, instead of memory within a
single machine, complicating the implementation of the system significantly.\\

\section {\label{sec:conclusion} Conclusions}
We describe the system architecture and  its implementation for the AARTFAAC all
sky  monitor,  an autonomous  and  real-time  image domain  transient  detection
machine piggy-backing  on the LOFAR radio  telescope.  The system consists  of a
diversity   of  heterogeneous   subsystems  ranging   from  FPGA   firmware,  to
heterogeneous  GPU machines,  with  final processing  carried  out on  commodity
computing machines.  Its aim is to  generate real-time triggers on the detection
of reliable transients, to enable their multi-wavelength followup.

Our implementation utilizes  a hierarchical routing of high bandwidth  data to a
central  correlator,  with co-processing  within  the  hierarchy to  spread  the
computing cost.  The most intensive computing  of the correlations of 1152 input
streams requires $\sim$ 40 TFLOPs, and  has been achieved on a system consisting
of 10 server grade GPU cards.  Our system operates in real-time, with an average
measured  latency of  1.1 seconds  to  generate calibrated  all-sky images.



\noindent  \textbf{Acknowledgments} This  work  is funded  by  the ERC  Advanced
Investigator  grant no.  247295 awarded  to Prof.   Ralph Wijers,  University of
Amsterdam.  We thank The Netherlands Foundation for Radio Astronomy (ASTRON) for
support provided in carrying out the commissioning observations.

\bibliographystyle{ws-jai}
\bibliography{ref}

\end{document}